\documentclass{interact}
\usepackage{graphicx}
\usepackage{csquotes}
\usepackage{url}
\usepackage{breakurl}
\usepackage{multirow}
\usepackage{multicol}

\usepackage{epstopdf}% To incorporate .eps illustrations using PDFLaTeX, etc.
\usepackage[caption=false]{subfig}
\usepackage[numbers,sort&compress]{natbib}% Citation support using natbib.sty
\bibpunct[, ]{[}{]}{,}{n}{,}{,}% Citation support using natbib.sty
\renewcommand\bibfont{\fontsize{10}{12}\selectfont}% Bibliography support using natbib.sty
\makeatletter% @ becomes a letter
\def\NAT@def@citea{\def\@citea{\NAT@separator}}% Suppress spaces between citations using natbib.sty
\makeatother% @ becomes a symbol again

\theoremstyle{plain}% Theorem-like structures provided by amsthm.sty

\theoremstyle{definition}

\theoremstyle{remark}

\begin{document}

\title{Battery draining attacks against edge computing nodes in IoT networks%
\footnote{This is an Accepted Manuscript of an article to be published by Taylor \& Francis in Cyber-Physical Systems, available online: \url{https://www.tandfonline.com/10.1080/23335777.2020.1716268}.}}

\author{
\name{Ryan Smith, Daniel Palin, Philokypros P. Ioulianou*\thanks{*CONTACT Philokypros P.\ Ioulianou. Email: pi533@york.ac.uk}, Vassilios G.\ Vassilakis, and
        Siamak F.\ Shahandashti}
\affil{Dept. of Computer Science, University of York, York, United Kingdom}
}
\maketitle

\begin{abstract}
%As the use of smart devices starts to increase so does the risk for potential attacks. Although this is the case, there does not seem to be much of a rush from manufactures to attempt to improve security for their end users. As a result, smart devices remain vulnerable to many attacks and more research is needed to explore their effects. 
Many IoT devices, especially those deployed at the network edge have limited power resources. A number of attacks aim to exhaust these resources and drain the batteries of such edge nodes. 
In this work, we study the effects of a variety of battery draining attacks against edge nodes. Through simulation, we clarify the extent to which such attacks are able to increase the usage and hence waste the power resources of edge nodes. 
Specifically, we implement hello flooding, packet flooding, selective forwarding, rank attack, and versioning attack in ContikiOS and simulate them in the Cooja simulator, and measure and report a number of time and power resource usage metrics including CPU time, low power mode time, TX/RX time, and battery consumption. 
Besides, we test the stretch attack with three different batteries as an extreme scenario. 
Our extensive measurements enable us to compare the effectiveness of these attacks. 
Our results show that Versioning attack is the most severe attack in terms of draining the power resources of the network, followed by Packet Flooding and Hello Flood attacks. Furthermore, we confirm that Selective Forwarding and Rank attacks are not able to considerably increase the power resource usage in our scenarios. 
By quantifying the effects of these attacks, we demonstrate that under specific scenarios, Versioning attack can be three to four times as effective as Packet Flooding and Hello Flood attacks in wasting network resources, while Packet Flooding is generally comparable to Hello Flood in CPU and TX time usage increase but twice as powerful in draining device batteries. 
\end{abstract}

\begin{keywords}
Edge computing, Internet of things, smart sensors, battery draining attacks, Cooja, ContikiOS
\end{keywords}

Word count: 6941

\section{Introduction}
Information technology has grown throughout the past century to become one of the most prominent features in our lives. This can partially be credited to devices becoming smaller, cheaper and yet more powerful. These changes have caused certain devices to not just become prominent in business but also in people’s homes and their everyday life. With companies such as Amazon attempting to get smart speakers into the homes of millions and other companies trying to do the same with taheir smart devices, there seems to have been some neglect in making them as secure as possible. This may seem more serious when considering the predictions that by 2020 more than 50 billion devices will be deployed to serve the Internet of Things (IoT) \cite{kanan2016iot}. 

In recent years IoT has been becoming increasingly popular, and with the promise to make peoples' lives easier more people are using resource-constrained IoT devices. Such devices have the ability to quickly and accurately measure and analyze important information such as levels of air pollution, healthcare indicators and can even be used in military applications. However, with the number of users increasing so does the potential for cyberattacks. Companies' main goal is to release their devices as soon as possible. That means that few devices have regular firmware upgrades \cite{ioulianou2019battery} and they are usually paired with weak components making them increasingly appetizing for attackers. One of the main components that is particularly vulnerable is the small battery these devices often contain. Even though IoT is becoming a necessary part of everyday life, there are many serious vulnerabilities that come along with it, and until they are resolved the question will remain: Do we want simplicity or security?

In this paper, network edge devices, and more specifically resource-constrained IoT devices are examined. These devices will often contain batteries that should be easy to drain, given that the right attack is carried out. With IoT becoming increasingly popular, and with it becoming easier for home and businesses to adopt these devices to their needs, the threat of these types of attacks is becoming more worrying. This work studies the impacts of hello flooding, packet flooding, selective forwarding, rank attack, stretch attack and versioning attack by implementing them in ContikiOS and simulating them in Cooja simulator. These attacks focus on exhausting the batteries of the devices.

The key contributions in the paper are provided as follows:
\begin{itemize}
    \item Several simulation tools were studied and compared in terms of accuracy, and their ability to measure power consumption and simulate edge of network devices.
    \item Six well-known battery draining attacks are implemented in ContikiOS and then simulated in Cooja simulator.
    \item Effects on nodes' batteries for each simulated attack are studied. Several metrics were used to make conclusions on how the attacks exhaust batteries.
    \item A comparison among attacks is presented showing the comparative severity of the attack based on several energy and time consumption metrics.
\end{itemize}

The rest of the paper is organized as follows.
In Section \ref{background}, we present our considered Cisco's 7-layer IoT architecture, the technologies used in IoT devices and discuss the IoT security concerns. 
In Section \ref{attacks}, we briefly discuss the attacks that may occur in IoT networks. 
In Section \ref{related}, we present the most significant works that study the effects of battery draining attacks in IoT. 
In Section \ref{simulators}, we discuss the software components used in our work.
In Section \ref{implementation}, we describe the implemented attacks, assumptions, and simulated scenarios. 
In Section \ref{results}, we present our experimental results obtained using the Cooja simulator.
In Section \ref{comparison}, we compare and evaluate the results from different attacks.
In Section \ref{conclusion}, we conclude and discuss possible future research directions.

\section{Background} \label{background}
\subsection{IoT architecture} \label{architecture}
Defining IoT is a heavily debated topic as well as defining what an IoT architecture is. However, what has been agreed on is that for the concept of IoT to work it needs to consist of a network, sensor and communications \cite{gigli2011internet}. One of the most detailed architectures to be used in this paper is the 7-layer model proposed by Cisco \cite{cisco_2014}.
As shown in Figure \ref{ref:cisco}, it consists of the following layers:

\begin{figure}
\centerline{
\includegraphics[height=5cm,keepaspectratio,width=\linewidth]{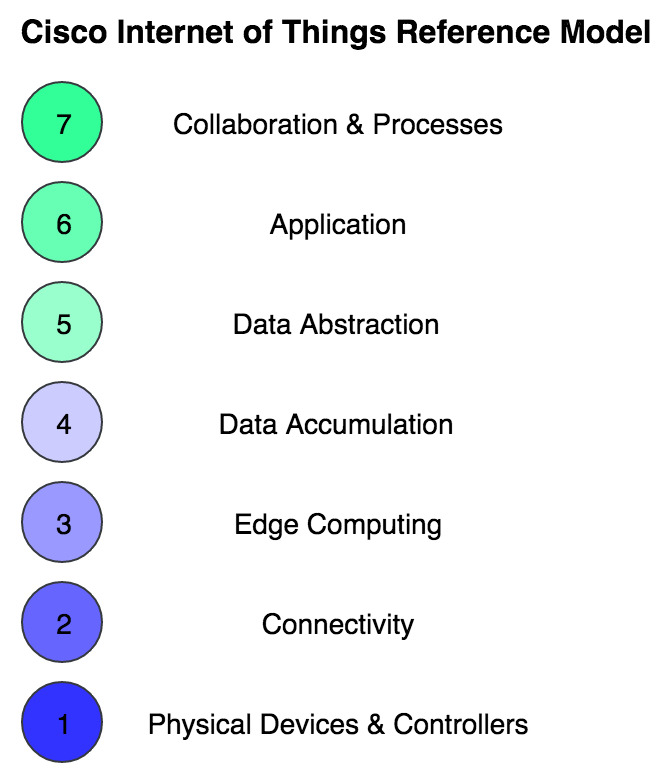}
}
\caption{7-layer IoT architecture proposed by Cisco}
\label{ref:cisco}
\end{figure}

\textit{Level 1 Physical Devices and Controllers} - This layer contains the things in the IoT. This includes a wide range of endpoint devices which can send or receive information (e.g., sensors and radio-frequency identification (RFID) readers) \cite{mosenia2017comprehensive}.

\textit{Level 2 Connectivity} - This level contains all components able to transmit information. The transmissions can be between devices in the first level, between the components in this level or between the first and third level.

\textit{Level 3 Edge (fog) Computing} - This is the first level where data processing occurs. Large amounts of information may be collected here. However, this level allows the system to only pass on relevant packets. This is essential as it reduces the load on higher levels. It also allows data to be formatted, expanded or decoded before it is needed by processing.

\textit{Level 4 Data Accumulation} - Until this point the data is in motion. Typically, the required information processing cannot be done at network speeds. As a result, this level converts the data in motion to data at rest. This means the data gets stored so that applications can access it when necessary. Data can also be transformed, recombined and recomputed ready for usage in the higher levels.

\textit{Level 5 Data Abstraction} - This level allows data to be stored in a more efficient way to improve performance of the higher levels. Some of the operations are normalisation, indexing, formatting, validation, consolidation of data and providing access to multiple data stores \cite{mosenia2017comprehensive}.

\textit{Level 6 Application} - The information is interpreted at this level and the applications interact with the data accumulation and data abstraction layers. The possible applications at this level can be very varied across different markets. For example, control applications and business intelligence applications are likely to use the data in very different ways.

\textit{Level 7 Collaboration and Processes} - The highest level pulls everything together and the system is useless unless the information provided at this step is useful. People should cooperate and use the data from the IoT system to make informed decisions.

\subsection{Technologies in IoT devices}
IoT started out with RFID and was seen as a way to get information about a product which was fitted with RFID chip. This attempt to get everyday objects into cyber space has carried on from there with various technologies which this section will explore.

One of the most used technologies in computers is Wi-Fi. With many companies having almost total Wi-Fi coverage it seems an almost obvious solution for IoT as well. However, it may not always be the most appropriate choice. This is because, as mentioned earlier, IoT devices tend to have small batteries and some Wi-Fi technologies, in particular 802.11a/b/g/n/ac for IoT, have a high energy usage \cite{jardosh2005understanding}. Although the benefits are clear, such as low cost and easy deployment, devices with smaller batteries may not see this technology as a feasible choice. There are, however, alternatives to this technology. 

Wi-Fi HaLow (802.11ah) \cite{Wi-FiAlliance} was designed to reduce power consumption by using wake/sleep periods and is a better choice for IoT. Furthermore, it has 1-kilometre range. However, this technology has never really taken off due to the fact that it requires a specialised access point and hardware. 

There is a new technology arriving called High-Efficiency Wireless (HEW) (802.11ax) \cite{deng2014ieee} which is being designed for IoT. This has the benefits of having a low power consumption and although it may not have the range of other technologies, for many devices it will be seen as very useful, especially for wearable technology. 

Bluetooth is seen as an inexpensive and low radius technology that can benefit some types of IoT networks. This technology has a small radius and is useful for devices such as printers and some PC’s. Older versions of Bluetooth have been somewhat inadequate for the usage of IoT, due to the low range and security concerns. However, newer Bluetooth technology, in particular Bluetooth 5 has higher data rate and higher radius than ever before at one kilometre, making Bluetooth a real contender for an effective IoT technology \cite{ray2016bluetooth}. It is also known that Bluetooth and Wi-Fi can co-exist, so it may be worth noting that if some devices require Wi-Fi and some require Bluetooth that they can still effectively work together. That means that using the right type of technology in a device can keep the power usage as low as possible, not just for the device but also for the network as a whole \cite{lansford2001wi}.

\subsection{The RPL protocol} \label{rplOver}
A new routing protocol for IoT devices is IPv6 Routing Protocol for Low-Power and Lossy Networks (RPL). According to \cite{rfc6550}, RPL is a standardised lightweight protocol that is mostly used in 6LoWPAN networks. By using an Objective Function (OF), RPL creates a Destination-Oriented Directed Acyclic Graph (DODAG) between the nodes in a 6LoWPAN network. OFs enhance routing metrics such as the Expected Transmission Count (ETX) in order to form routes in the DODAG.
Both unidirectional traffic towards the DODAG root and bidirectional traffic between nodes and the root are supported by the protocol. A single 6LoWPAN network can have more than one RPL instances, and a global DODAG can have a local RPL DODAG among several nodes.
The IPv6 address of the node is used as its ID. Nodes also store their DODAG neighbours in a list and they can have one or more parents, except for the root. In addition, all nodes have a rank where it's lowest at the root.

RPL comes with new ICMPv6 control messages. DODAG Information Object (DIO) messages are initially sent by the root. These messages contain information about the rank of the broadcasting node (which is the distance of the node from the backbone network), the OF, and the DODAG ID. Apart from that, DIO messages help maintaining the DODAG. If a node gets a DIO message, it determines its rank (according to the advertised rank in the received message) and the cost of getting to the sending node from itself. Each node sends these messages in intervals based on trickle timer \cite{trickle1}. This timer also prevents sending unnecessary DIO messages.

In order for a node to join the network, it must get a DIO message or multicast a DODAG Information Solicitation (DIS) message to request a DIO message. When other devices get the DIS message, they will start broadcasting DIO messages, and the new node can join the DODAG.
Then, a Destination Advertisement Object (DAO) message is sent by the new node to its parent. In some cases, parents may send DIO messages to sub-DODAG in order to request DAO messages.
DAO messages are important for creating downward routes (from root to node). 
Nodes update their routing table when a DAO message is received. If routing tables are empty or if packets are destined for the root, the node will forward a packet up to its most preferred parent.

\subsection{Security concerns of IoT devices}
There are multiple concerns regarding security when it comes to IoT devices; because their size puts constraints on what they can feasibly achieve to keep them secure. This section will focus on the security issues of Level 1 of the IoT architecture presented in Subsection \ref{architecture}. At this level, vulnerabilities of physical devices and controllers are discussed. Although higher layers tend to have a high level of security, the devices themselves that people interact with tend to have very little security, in an effort to keep them small and low powered.

\subsubsection{Processing power}
Many IoT devices have the ability to provide on-device processing. This includes converting data to another format, enhancing data so the most important components are highlighted and making sure data hits a set of rules by validating it \cite{google}. For example, smart sensors usually need wireless connectivity and other components which require processing power. This is difficult to achieve in a small device. Thus, some of these devices will have a power consumption as little as 50 to 100 Million Instructions per Second (MIPS) \cite{voica_2016}. Due to cost restrictions, it is not practical to deploy significant resources to help improve the security of these devices. An important work that looks at creating a secure low powered IoT system is from Brooks et al. \cite{brooks2017ultra}. The authors investigate the possibility of building a low cost and low-powered processor yet to still ensuring its high security. %One way to achieve this is by including physically unclonable functions (PUFs) that rely on device properties to provide a unique signature that provides an authentication code for a given system.

\subsubsection{Battery life}
One of the biggest concerns regarding IoT is the size of the battery, and how easily the energy consumption of a device can be manipulated to quickly drain the battery. A report from Silicon Labs \cite{batterysize} looks at how people expect to have more functionality for their devices, without increasing their size or cost. This quite often leads the battery to remain inadequate and vulnerable. There are very few options that can be physically used in IoT devices due to size constraints. The main options are AAA, CR2032, CR123A, and CR2 batteries. Table \ref{tab:batteries} shows the differences between these batteries.

\begin{table}
    \centering
    \tbl{Different types of commonly used IoT batteries}    
    {\begin{tabular}{@{}lcccc@{}}
    \toprule 
         \textbf{Name} &
         2x AAA &
         CR2032 &
         CR123A &
         CR2 \\
         \midrule
         \textbf{Voltage} & 
         3V &
         3V &
         3V &
         3V \\
         \midrule
         \textbf{Capacity} &
         1000 mAh &
         225 mAh &
         1500 mAh &
         800 mAh \\
         \midrule
         \textbf{Material} &
         Alkaline &
         LiMnO2 &
         Lithium &
         Lithium\\
         \bottomrule
    \end{tabular}}
    \label{tab:batteries}
\end{table}

In an effort to solve the battery issues, Calhoun et al. \cite{calhoun2015ultra} proposed using a battery-free technology for IoT devices. Specifically, they suggest using ambient radio signals to power devices. However, this has not been seen as feasible, with more money being spent in improving current batteries, rather than going battery-free.

\subsubsection{Bandwidth}
The amount of data produced by IoT devices at Level 1 is so large that it is almost impossible for it to transmit directly to the cloud servers through the gateway without any compression. This is because when the device attempts to transmit such a large amount of uncompressed data it will cause packets to be dropped and lost. This means that it is necessary for the data sent to be pre-processed, kept of a high quality and made so that it does not use up too many resources \cite{zhao2013survey}. For IoT nodes, the Transport Layer Security (TLS) protocol is typically used to create a secure connection between the different layers when transporting data. Yet, this does not keep these devices or their data secure. This is because the device itself has not been authenticated and these encryption methods do nothing to prevent an attacker taking over the node \cite{secintel}.

\section{Battery draining attacks in IoT} \label{attacks}
When designing an IoT device, one of the main overlooked issues is the size of the battery that is being used. This is often due to various reasons such as keeping the device small and portable. Due to this small size, devices are vulnerable.  These issues become more serious when, for example, looking at types of devices that patients may wear to detect changes in their body. Below we explore specific attacks that are designed to drain the batteries of IoT devices.

\subsection{Denial of sleep}
Various attacks can cause smart devices to consume more energy. Krentz et al. in \cite{krentz2017countering} explore three denial of sleep attacks. The first type of attack is called `Ding Dong Ditching'. This type of attack uses a combination of four denial of sleep attacks to try to attack a device. This increases the potential for at least one of the attacks to work and to get a positive result. The first type of attack is a jamming attack. A jamming attack causes the network to become unusable by people and can be accomplished by emitting RF signal \cite{raymond2009effects}. This type of attack is called a constant jammer where the MAC layer is bypassed by constantly transmitting random bits of data through an RF signal. By constantly transmitting data the transfer does not wait for a channel to become idle and therefore does not follow the MAC layer rules \cite{thakur2013introduction}. 

The next part of this attack is broadcasting, where an adversary will send a ping to every host, which in turn will send an ICMP response. Using this information, the adversary can carry out further attacks that include the other nodes on the network being your slaves.

The final aspect of this attack is a droplet attack. This occurs when the beginning of an 802.15.4 frame is sent and then it stops transmitting. This leaves the receiver of this frame in receive mode. By using all of these attacks or just a combination you can create a good attack where elements of the attack are likely to get through and help drain the batteries of the edge devices. 

\subsection{Flooding attack}
The main type of flooding attack is a `Hello' Flooding attack. This is where a malicious RPL node creates massive amount of traffic by sending DIS messages to other RPL nodes, causing the recipient nodes to respond by sending DIO messages. As a result, congestion is created in the network and nodes are energy exhausted. The other element to this attack is that if the attacker has a large enough broadcasting range they can make nodes think that they are their neighbour and forward them packets. They can also make the legitimate node try to send them packets thinking they are next to them. However, packets will end up getting lost in the network causing it to flooding and slow down more because node are actually far away \cite{singh2010hello}.

\subsection{Vampire attack}
This type of attack will put a constant strain on device's small batteries causing them to stop working. Two types of vampire attacks are typically distinguished \cite{vasserman2013vampire}. The first is a stretch attack and the second is a carousel attack. A stretch attack sends a packet along a longer route around the network by changing the information in the header of the packet. This attack could be paired with certain other attacks to make them more powerful and effective. Authors in \cite{umakanth2013detection} have shown that using a stretch attack whilst not being paired can increase energy consumption by a factor of 4. Moreover, this attack can be easily carried out, especially when the network does not employ authentication, as the attacker can simply change the route of the packet in the header.

Regarding carousel attack, it loops a packet around a network before it eventually arrives at its location. Both of these attacks can be used in combination with other types of attack. However, if used alone they may be left undetected making it difficult to defend against.

There have been multiple attempts to prevent these attacks. However, the defence mechanisms run into the issue of using more power than the attack itself. One solution that has been looked at is a zero-power notification which will create an alert when an attack is being detected \cite{rathi2017wearable}.

\section{Related work} \label{related}
Authors in \cite{krentz2017countering} looked at `Ding-dong ditching' and stated that in order to find out the energy consumption of these attacks one needs to look at all the aspects separately. Firstly, they tested jamming attacks. To do this, they sent out an emitted radio noise in four subsequent runs and the attacked node had no defences set up to protect itself. This attack showed that after 4.286 milliseconds the node had a mean energy consumption of 0.372 mJ. Once again, running a broadcast attack shows that it had a mean energy consumption of 0.7 mJ. Next, they tested unicast attack and this showed a mean usage of 0.7 mJ again. Authors also looked at some ways to defend against these attacks. They suggested that by using an idea called dozing and another called POTR, one can significantly reduce the amount of power consumption from these attacks.

In another work, Vasserman et al. \cite{vasserman2013vampire} studied the effects of stretch and carousel attacks. Firstly, a legitimate packet was sent from source to sink. Authors recorded the average power consumption which was around 0.008 as a fraction of node energy consumed. Then, they deployed a stretch attack and measured again the power consumption throughout the network. They found an average power consumption of 0.01 as a fraction of node energy consumed. Finally, they deployed carousel attack which had an average power consumption of 0.07 as a fraction of node energy consumed. Authors showed that when using a carousel attack the power consumption can be 8.75 times higher than when a legitimate node is being sent around. The results of power consumption were based on number of packets sent from one randomly placed malicious node.

Another interesting work in the field of RPL attacks is from Mayzaud et al. \cite{mayzaud2014study}. Authors investigate how the RPL versioning system can be exploited to gain an advantage in the topology and also force children nodes to route packets via a malicious node. They use several metrics such as overhead, delivery ratio, end-to-end delay, rank inconsistencies and loops. However, they do not look at the energy utilised by nodes during the attack.

In \cite{wallgren2013routing}, authors are studying the RPL protocol and how it can be exploited. They implement well-known routing attacks in ContikiOS using Cooja simulator. After simulating attacks, they propose an Intrusion Detection System (IDS) which implements a lightweight heartbeat protocol. Although their work looks promising, they do not present results of nodes' energy consumption during several routing attacks.

All in all, it is clear that certain aspects of edge node attacks have not been sufficiently covered. Existing works describe how attacks increase energy consumption of devices, but few explore and quantify to what extent exactly established Denial-of-Service (DoS) attacks will affect battery consumption, especially in comparison with other types of attack and in ad hoc scenarios. 

\subsection{Research questions}
The overall goal of our paper is to examine the negative effects of the most known DoS attacks on IoT devices. The focus is on the battery and on finding out the most severe attack that could occur in a smart device. The following research questions will be discussed:
\begin{enumerate}
    \item How easy is to implement the various DoS attacks in ContikiOS? Which simulator is the most accurate and allows measuring of device's power consumption?
    \item What are the consequences on battery life for each of the implemented attack? How do the attacks compare in terms of quantitative metrics of time and power consumption? 
    \item How effective is the Stretch Attack in different scenarios in which devices use different battery types and sizes? 
\end{enumerate}

\section{Simulation tool} \label{simulators}
 Several criteria were considered for choosing the simulation tool, first and foremost is accuracy. The simulation tool has to be well established and have a common consensus that the results it could produce are accurate. Secondly, it is the ability to measure power consumption. Many network simulation tools do not have the ability to measure power. However, a few have additional modules can be used to do this. Finally, it should be able to simulate network edge topologies. Considering that the purpose of this experiment was to look at smart devices, this criterion is also important. The network has not only had the attack packets flowing through the network but also legitimate data, as this could also affect the power consumption. Table \ref{tab:simulators} lists the potential network simulation tools and provides a short description about whether or not they are suitable regarding each of the three criteria.

\begin{table*}[ht]
    \centering
    \tbl{Potential tools for IoT simulations}
    {\begin{tabular}{@{}llll@{}}
    \toprule
         \textbf{Simulation tool} & 
         \textbf{Accuracy} &
         \begin{tabular}{@{}l@{}}\textbf{Power}\\\textbf{measurement}\end{tabular}  &
         \begin{tabular}{@{}l@{}}\textbf{Edge of network} \\\textbf{simulation}\end{tabular}  \\
         \midrule
         TOSSIM  & 
         \begin{tabular}{@{}l@{}}Commonly used in\\literature to accurately\\simulate `TinyOS'.\end{tabular} &
         \begin{tabular}{@{}l@{}}Add on through\\Power Tossim.\end{tabular} &
         \begin{tabular}{@{}l@{}}Yes. Designed to\\run using `TinyOS'.\end{tabular} \\
         \midrule
         NS-2/3 &
        \begin{tabular}{@{}l@{}}Testing done to\\compare it to real\\world situations.\end{tabular} &
         No. &
        \begin{tabular}{@{}l@{}}Yes. Needs\\manual configuration.\end{tabular} \\
        \midrule
        Cooja &
        \begin{tabular}{@{}l@{}}Testing done to\\compare it to real\\world situations.\end{tabular} &
        \begin{tabular}{@{}l@{}}Add on through\\energest and powertrace.\end{tabular} &
        \begin{tabular}{@{}l@{}}Designed to be used with\\ wireless sensor networks.\end{tabular}\\
        \bottomrule
    \end{tabular}}
     \label{tab:simulators}
\end{table*}

Although our initial simulations were done using NS-2 and then swapping to NS-3, it was decided that these simulators may not be suitable due to not being accurate in measuring power consumption \cite{lahmar2012wireless}. Therefore, the network simulation tool was chosen to be Cooja which runs on ContikiOS \cite{contiki}. An advantage of using Cooja is that it enables uploading of simulated firmware into real physical sensors.

\section{Implementing DoS attacks in Cooja} \label{implementation}
Our first goal is to explore how various types of DoS attacks can be used to drain the batteries of smart devices. For this reason, several attacks were implemented and are presented in this section.

\begin{figure}[ht]
    \centerline{
    \includegraphics[scale=0.3]{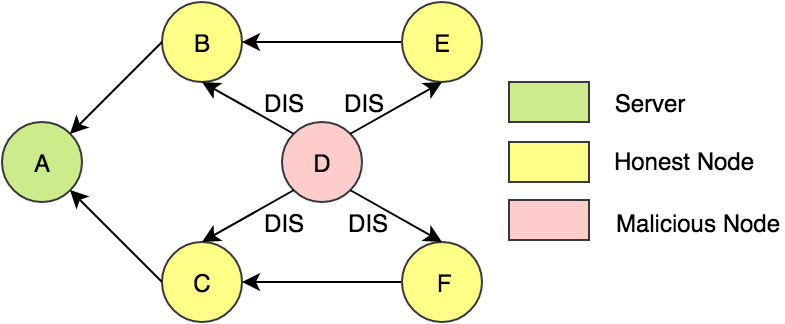}
    }
    \caption{Hello flooding}
    \label{fig:hellofloodtop}
\end{figure}

The first attack to be explored is the `Hello' flooding attack. This is a Layer 2 attack and, therefore, will affect device data transmission. As mentioned before, when a new node joins a network, it sends DIS messages to all its neighbours. These neighbours then respond with DIO messages. In this attack, a malicious node sends too many DIS messages very frequently causing network congestion.
The attack diagram is given in Figure \ref{fig:hellofloodtop}. It shows a malicious node forwarding DIS messages (Hello packets) to nodes within the network.

A similar implemented attack is the packet flooding. This attack occurs when a malicious node rapidly sends packets to other nodes in the network forcing them to run their checking mechanisms \cite{mosenia2017comprehensive}. As a result, the energy of the nodes is exhausted. In our work, malicious node sends unicast packets to server in order to exhaust it. 

%\begin{figure}[htp]
%    \centering
%    \includegraphics[scale=0.25]{packetflood.png}
%    \caption{Packet flood attack topology}
%    \label{fig:packetfloodatt}
%\end{figure}

\begin{figure}
    \centerline{
    \includegraphics[scale=0.3]{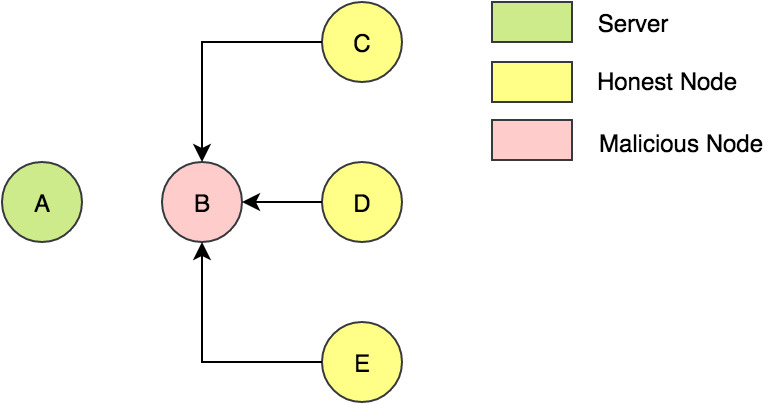}
    }
    \caption{Blackhole attack}
    \label{fig:selfwd}
\end{figure}

Another attack that we consider is the selective forwarding. This kind of attack occurs when a malicious node chooses which packets to forward to the next node. It is called a greyhole attack if some packets pass through and some are dropped. If all packets are dropped, it is called a blackhole attack \cite{jan2013denial}. Figure \ref{fig:selfwd} shows the diagram of a blackhole attack. Nodes C, D and E send packets to node A via a malicious node B. Yet, B drops all packets and nothing is received at node A. This attack aims to disrupt routing paths. Our implementation is a greyhole attack since the DIO messages are allowed to pass from malicious node while the data packets are dropped. This can be modified to drop different ratios of packets or look at the contents of packets and drop important ones.
%Stretch attack. The reason for choosing this attack is that it seems like a natural progression to attempt to increase the potential of a Hello Flood attack. This is because rather than sending from the source to the sink, it uses a longer route to cause a higher rate of energy usage throughout the network. Therefore, an increase on the battery consumption of the network is expected. 

\begin{figure}[h!]
    \centerline{
    \includegraphics[scale=0.7]{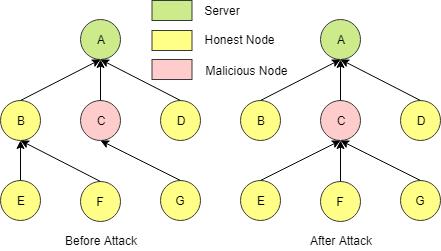}
    }
    \caption{Rank attack}
    \label{fig:rankatttop}
\end{figure}

The fourth implemented attack is the versioning attack. The aim of the attack is to again deplete the energy of the nodes. A malicious node increments the DODAG version number which is now inconsistent with other nodes in the network. RPL detects this error in version numbers and performs a global repair which unnecessarily rebuilds the network \cite{d'hondt_bahmad_vanhee_2016}.

The rank attack is the fifth implemented attack in ContikiOS. Attacker node advertises a low rank so that it attracts all nodes in the DODAG to connect to it so that all packets pass through it \cite{d'hondt_bahmad_vanhee_2016}. As a consequence, a malicious node exploits the RPL protocol to cause a sinkhole attack. This can be achieved if the rank of a malicious node is always one more than the root node. On the left of Figure \ref{fig:rankatttop}, the network diagram without the attack is presented. Here, all children are connected to the closest parents. On the right hand side, the diagram shows the network when the attack is initiated. All children choose malicious node C as their parent.

\section{Environment setup and scenarios}
\subsection{Contiki configuration}
In order to run experiments, we configured Cooja and ContikiOS as described in \cite{contiki}.
Moreover, Powertrace was setup and added to our project. This module is provided by ContikiOS and is able to track all packets and their transmission and receive rates. Also, the Zolertia Z1 mote was used as hardware platform which uses the MSP430 microcontroller and the CC2420 radio transceiver. Table \ref{tab:datasheet} presents the current consumption of each of Z1's components. A voltage of 3V for all the components has been used as it is the voltage supplied by two AA batteries and it is recommended by Z1 datasheet \cite{zolertiaz1}. Each of these batteries have a capacity ranging from 450 mAh to 2640 mAh depending on drain rate \cite{alkaline_handbook}.

\begin{table}[ht]
    \centering
    \tbl{Z1 component datasheet values}
    {  \begin{tabular}{@{}ccc@{}}
    \toprule
         \textbf{Mode} &
         \textbf{Voltage} &
         \textbf{Current Consumption} \\
        \midrule
        Power down &
        3V & 20 $\mu$ A \\
        \midrule
        IDLE & 3V & 426 $\mu$ A \\
        \midrule
        RX & 3V & 18.8 mA \\
        \midrule
        TX & 3V & 17.4 mA \\
    \bottomrule 
    \end{tabular}}
    \label{tab:datasheet}
\end{table}

\subsection{Assumptions and scenarios}
The purpose of our experiments is to explore how various attacks can be used to drain the batteries of network edge IoT devices. The experiments involve the attacks described in Section \ref{implementation}. These attacks are known to slow down a network but not much is known about the effects they have on battery-powered devices.

After deciding the simulator for our experiments, several other assumptions needed to be made. Firstly, the server and malicious node in all scenarios have infinite power. Secondly, the network for all the experiments should be the same so that each experiment can be set up accurately each time. This should allow the results to be more understandable.

\begin{figure}[h]
    \centering
    \includegraphics[scale=0.6]{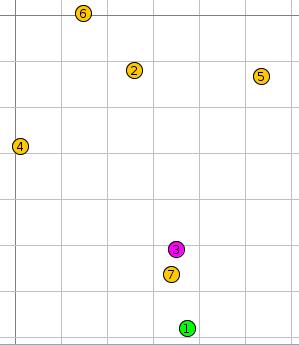}
    \caption{Simulated network}
    \label{fig:simnet}
\end{figure}

We considered a network of size seven with one server, five honest nodes, and one malicious node. We planned to study the effects of attacks when nodes are placed in an ad hoc manner, hence we generated a topology with random placements of nodes in Cooja. This topology is shown in Figure \ref{fig:simnet}, in which the server is shown in green, honest nodes in yellow, and the malicious node in purple. This configuration is simulated for 5 hours having only honest nodes and then the simulation was repeated after adding the malicious node. The malicious node runs a different attack in each scenario so that the effects of each attack are studied.
All the attacks were simulated using the same network topology. Therefore, the standard network size remained at seven, the standard battery size remained at two AA batteries and the number of compromised nodes remained at one.

\section{Experimental results} \label{results}
This section presents the results from simulating the attacks described in previous sections. The metrics that we consider are the following: 1) Total CPU time, 2) Total Low Power Mode (LPM) time, 3) Total TX time, and 4) Total RX time. The total time was averaged over all honest nodes. Results are shown in Figure~\ref{fig:usage}. The blue line in the graphs represents the resource usage when there is no attack (labelled ``Honest Network'' in subfigures).

\subsection{Packet flooding}
\begin{figure}[t]
\centering
\subfloat[Total CPU time\label{fig:cpu}]{%
\resizebox*{7cm}{!}{\includegraphics{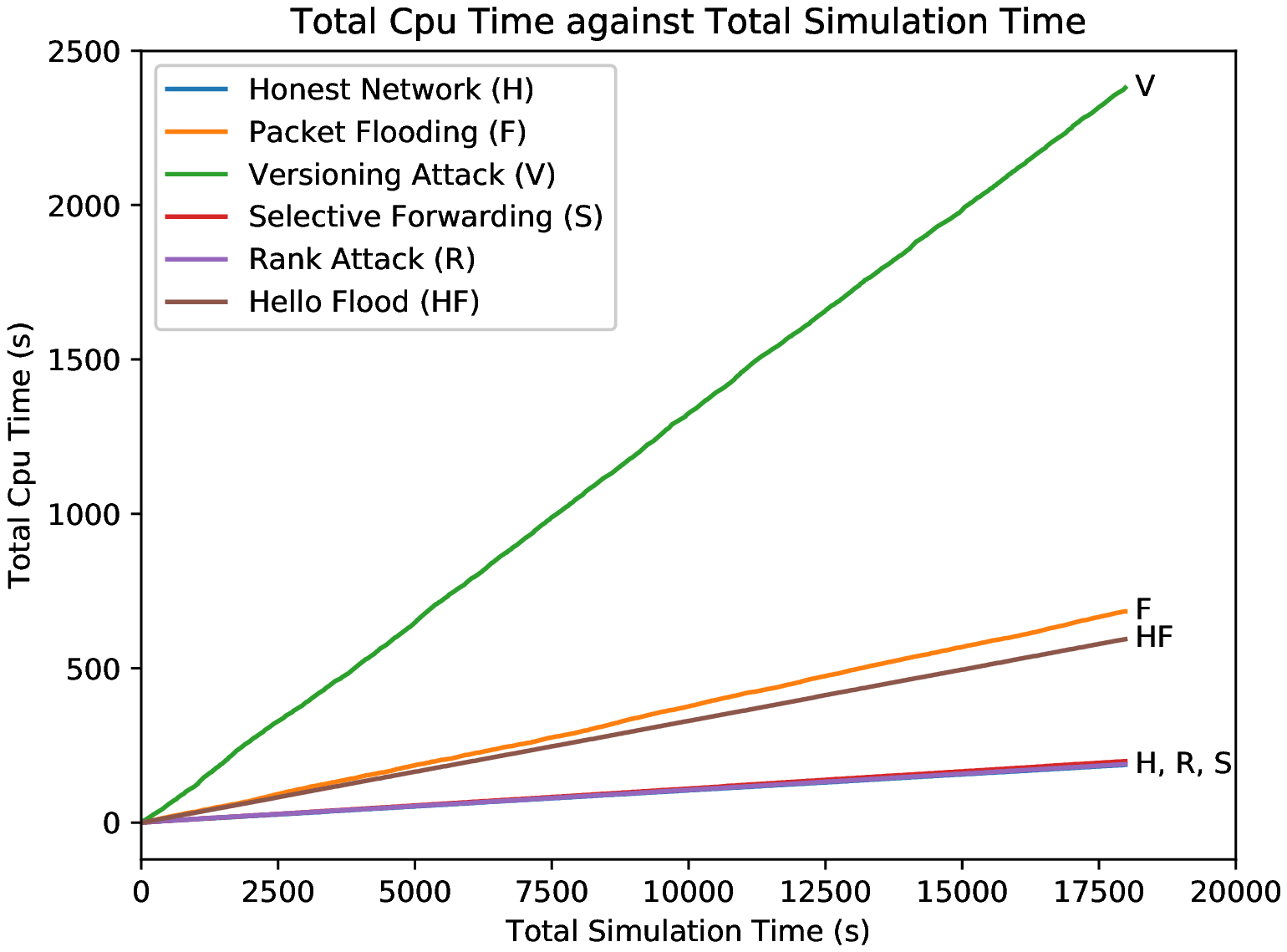}}}\hspace{5pt} 
\subfloat[Total Low Power mode time\label{fig:lpm}]{%
\resizebox*{7cm}{!}{\includegraphics{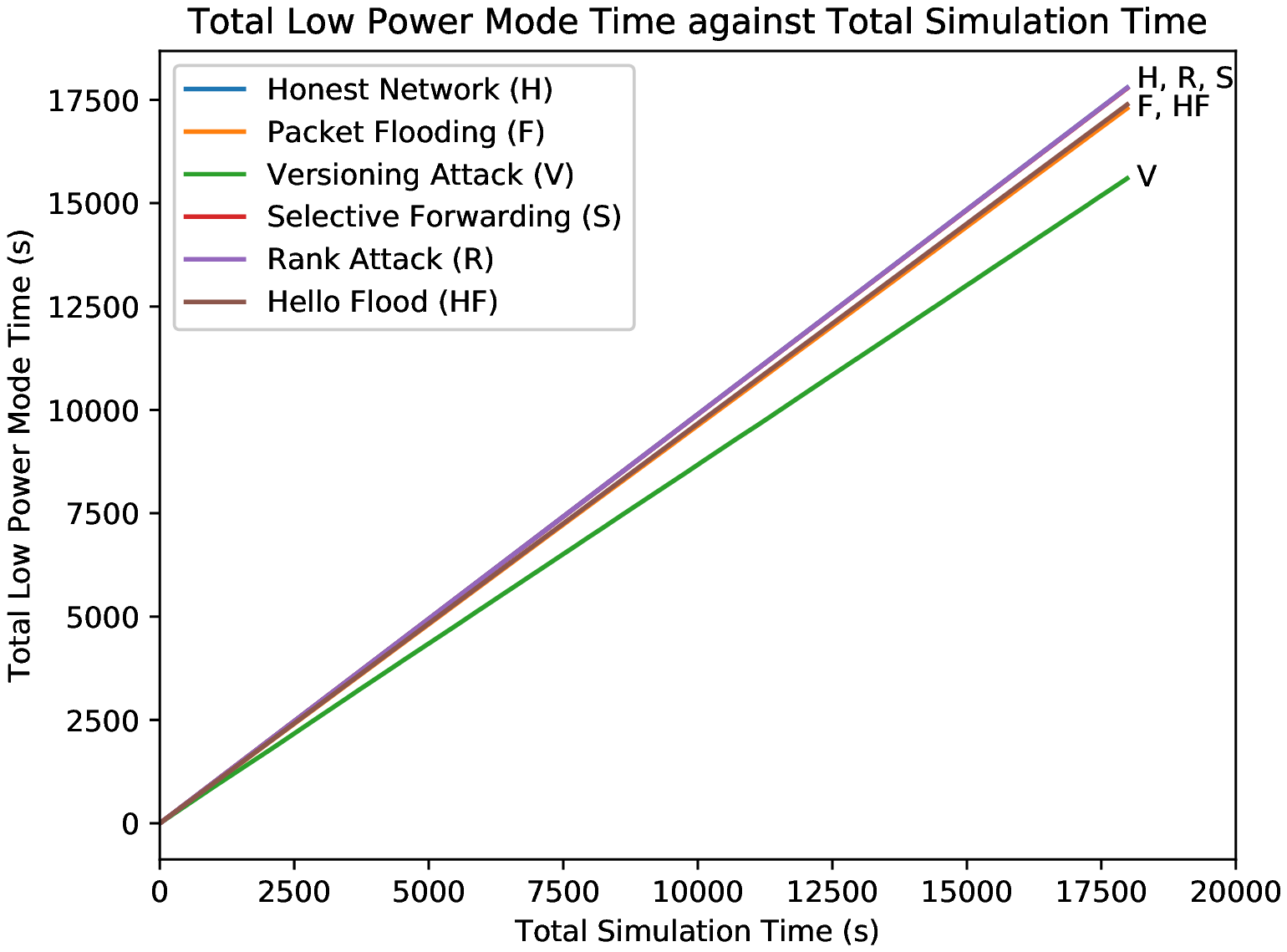}}}\vspace{5pt}
\subfloat[Total TX time\label{fig:txtime}]{%
\resizebox*{7cm}{!}{\includegraphics{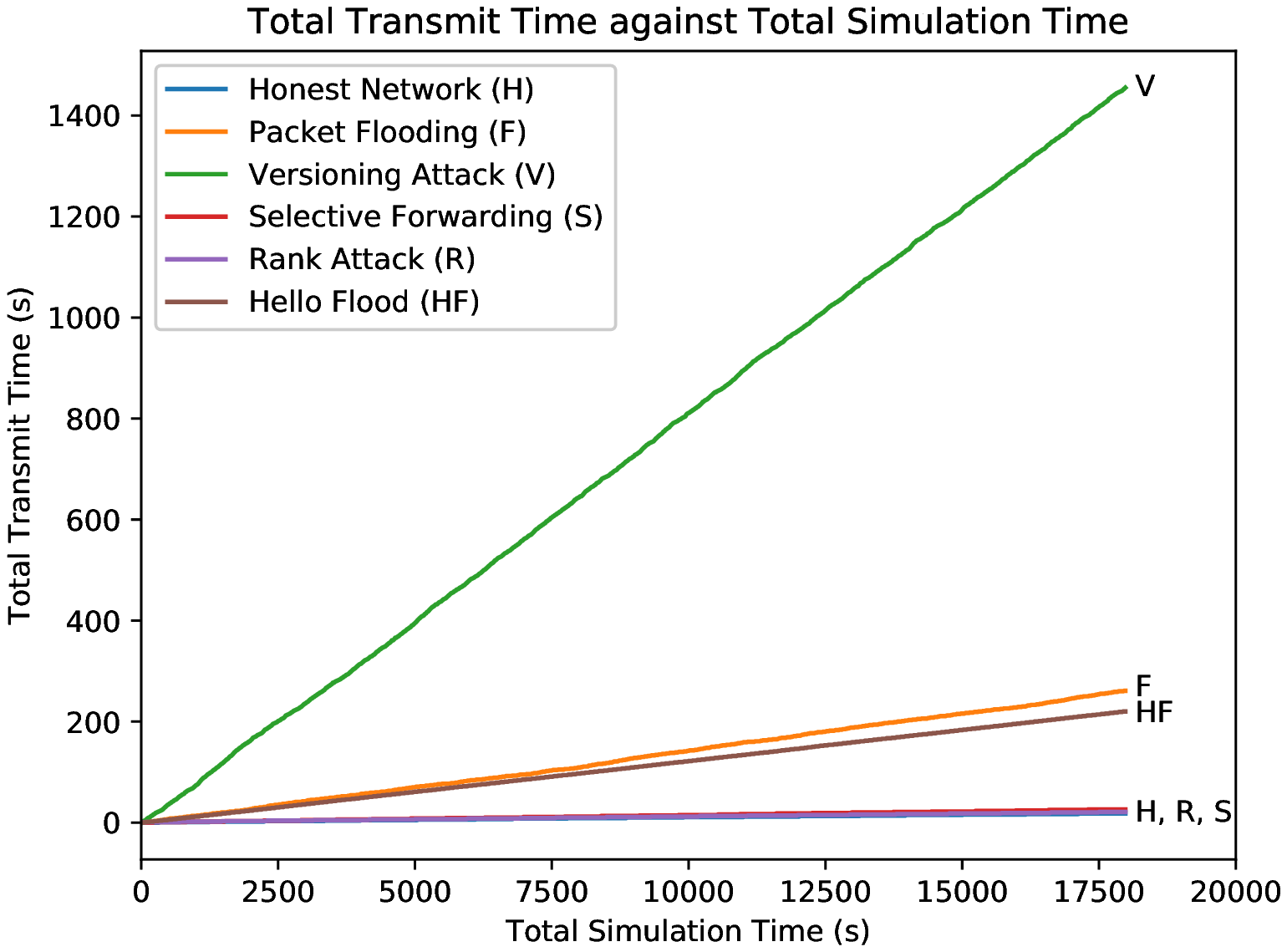}}}\hspace{5pt}
\subfloat[Total RX time\label{fig:rxtime}]{%
\resizebox*{7cm}{!}{\includegraphics{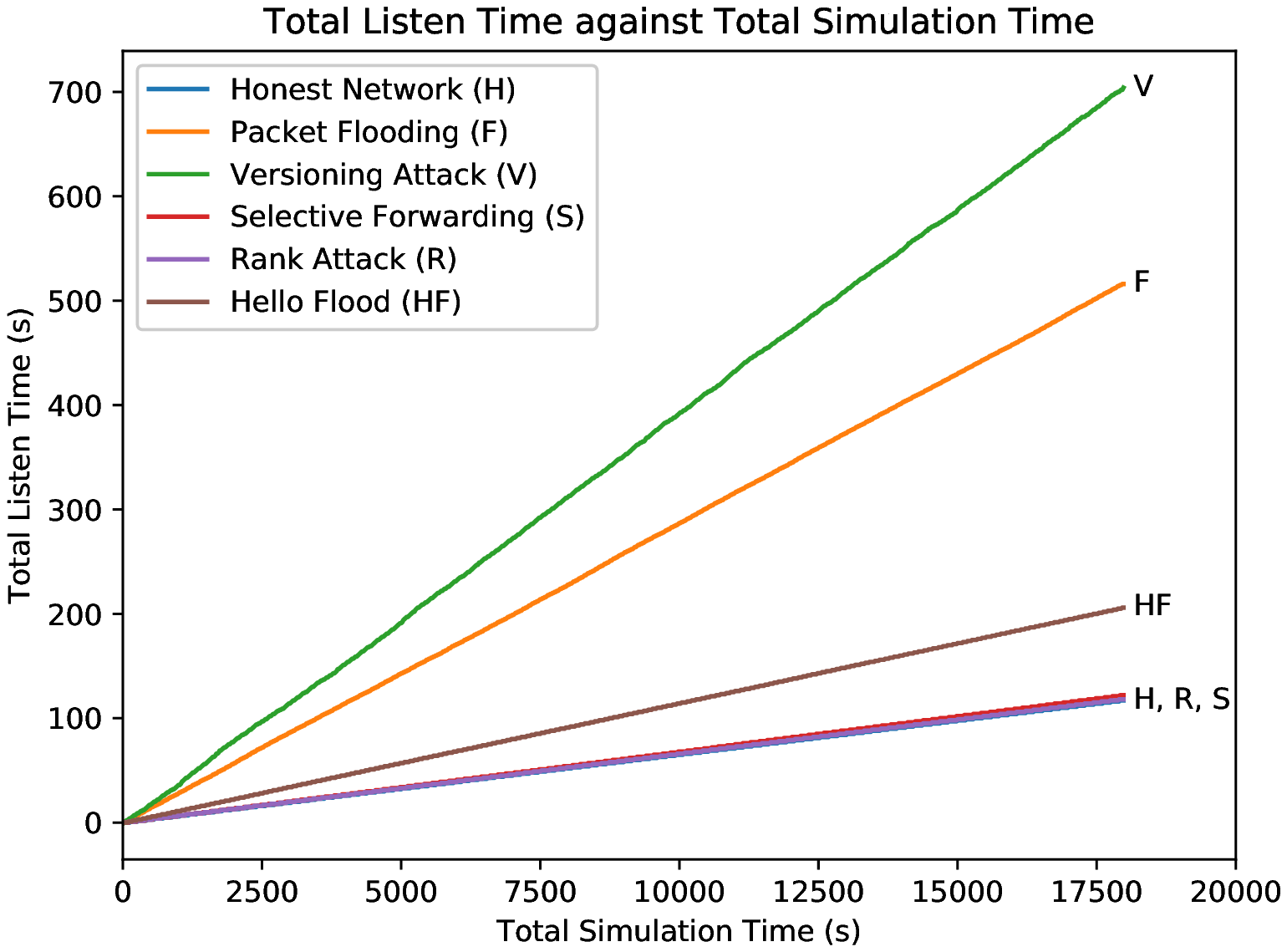}}}
\caption{Total resource usage over simulation time for various attacks} 
\label{fig:usage}
\end{figure}

As shown in Figure \ref{fig:cpu}, total CPU usage was increased by 265\% during the attack while nodes were in LPM for 3\% less of the time (Fig. \ref{fig:lpm}). This is expected as nodes are either in CPU or LPM, the time lost in LPM was gained in CPU mode. The time spent in TX mode by nodes is shown in Figure \ref{fig:txtime}. In the attack scenario, TX mode is increased by 940\%. Similarly, the time that nodes were in RX mode (Fig. \ref{fig:rxtime}) increased by 329\%. As a result, packet flooding affects mostly the time that nodes are in TX mode.

\subsection{Hello flooding}
We observe that the attack increased the CPU usage by 226\%  (Fig.~\ref{fig:cpu}) and that the nodes are in LPM mode for 2\% less of the time during the attack (Fig.~\ref{fig:lpm}). Again, as the node is either in CPU or LPM mode, the time lost in LPM was gained in CPU mode. Total TX time is increased by 780\% when attack happens as depicted in Figure \ref{fig:txtime}. Also as Figure \ref{fig:rxtime} shows, the time the node was in RX mode is increased by 81\% during the attack. Similarly to packet flooding, Hello flooding affects mostly the time in TX mode.

%\begin{figure}[h]
%\centering
%\subfloat[Total CPU time]{%
%\resizebox*{7cm}{!}{\includegraphics{hellocpu.jpeg}}}\hspace{5pt}
%\subfloat[Total Low Power mode time]{%
%\resizebox*{7cm}{!}{\includegraphics{hellolpm.jpeg}}}\vspace{5pt}
%\subfloat[Total TX time\label{fig:hellotxtime}]{%
%\resizebox*{7cm}{!}{\includegraphics{hellotx.jpeg}}}\hspace{5pt}
%\subfloat[Total RX time\label{fig:hellorxtime}]{%
%\resizebox*{7cm}{!}{\includegraphics{hellorx.jpeg}}}
%\caption{Hello flood:Total resource usage over simulation time} 
%\label{fig:hellores}
%\end{figure}

\subsection{Selective forwarding}
In the selective forwarding attack, all nodes send packets to the server through node 3 because it is in their range except node 7 which can directly reach the server. In this way, malicious node 3 can drop packets and forward only selected ones. Results in Figure \ref{fig:usage} show that usage figures of all the four considered resources are almost identical to those of the no-attack scenario. Hence, this attack has no considerable effect on the power usage of nodes but affects only the number of received packets of the server.

%\begin{figure}[h]
%\centering
%\subfloat[Total TX time]{%
%\resizebox*{7cm}{!}{\includegraphics{greytx.jpeg}}}\hspace{5pt}
%\subfloat[Total RX time]{%
%\resizebox*{7cm}{!}{\includegraphics{grey1.jpeg}}}
%\caption{Selective forwarding: Total resource usage over simulation time}
%\label{fig:greyusagepacket}
%\end{figure}

\begin{figure}[t]
    \centering
    \includegraphics[scale=0.6]{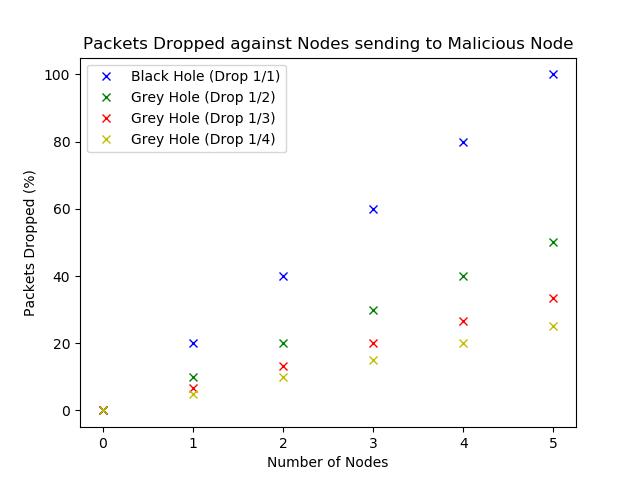}
    \caption{Selective forwarding: Percentage of packets dropped}
    \label{fig:packetdrop}
\end{figure}

The expected number of packets to be dropped during the simulation are illustrated in Figure \ref{fig:packetdrop}. Specifically, the graph shows the expected packet dropped ratio depending on the number of nodes sending to the malicious node and the drop ratio of the malicious node. In our experiment, 4 nodes are sending packets to the malicious node and all data packets are dropped. If we look at the blackhole points (blue) on the graph, when the nodes are 4 it shows that 80\% of all packets in the network should have been dropped. That is correct in our case as the server receives packets from node 7 only. The other colours represent different cases of greyhole attacks and the number of packets that should be dropped. Overall, the results of this set of simulations are as expected, i.e. the number of dropped packets grow linearly with the drop rate and with the number of nodes sending to the malicious node. We nevertheless include the results for completeness. 

\subsection{Versioning attack}
Figure \ref{fig:cpu} shows the CPU usage is increased by 1146\% during the attack. Moreover, Figure \ref{fig:lpm} depicts that nodes are in LPM for 12\% less of the time. Thus, during the attack the time lost in LPM was gained in CPU mode. Figure \ref{fig:txtime} represents the time that TX mode is activated in nodes. This has risen to 5720\% while RX mode is increased by 478\% as shown in Figure \ref{fig:rxtime}. That means that versioning attack has a significant effect on the network as a whole and a large impact on each of the nodes.

%\begin{figure}[h!]
%\centering
%\subfloat[Total CPU time\label{fig:vercpu}]{%
%\resizebox*{7cm}{!}{\includegraphics{vercpu.jpeg}}}\hspace{5pt}
%\subfloat[Total Low Power mode time\label{fig:verlpm}]{%
%\resizebox*{7cm}{!}{\includegraphics{verlpm.jpeg}}}\vspace{5pt}
%\subfloat[Total TX time\label{fig:vertx}]{%
%\resizebox*{7cm}{!}{\includegraphics{vertx.jpeg}}}\hspace{5pt}
%\subfloat[Total RX time\label{fig:verlx}]{%
%\resizebox*{7cm}{!}{\includegraphics{verlx.jpeg}}}
%\caption{Versioning attack: Total resource usage over simulation time} 
%\label{fig:veratt}
%\end{figure}

\subsection{Rank attack}
The results illustrate that the resources used for TX and RX are almost identical to those of the no-attack scenario. This is also the case for CPU and LPM modes. That means the attack alone has no effect on the power usage. It is mainly used to increase the impact of other attacks. For example, a blackhole attack could be combined with a rank attack to make it more harmful. Apart from that, this attack allows the malicious node to eavesdrop on all packets being sent to it. Although it does not affect the power usage, it can cause serious damage to an IoT network.

%\begin{figure}[h]
%\centering
%\subfloat[Total TX time]{%
%\resizebox*{7cm}{!}{\includegraphics{ranktrans.jpeg}}}\hspace{5pt}
%\subfloat[Total RX time]{%
%\resizebox*{7cm}{!}{\includegraphics{rankrx.jpeg}}}
%\caption{Rank attack: Total resource usage over simulation time} 
%\label{fig:rankatt}
%\end{figure}

%Figure showing stretch attack topology. Let's hide it to reduce figures.
%\begin{figure}[h!]
%    \centering
%    \includegraphics[scale=0.5]{mednet.jpeg}
%    \caption{Stretch attack topology}
%    \label{fig:midnet}
%\end{figure}

\subsection{Stretch attack}
Although the worst attack seems to be versioning attack, we were still interested in implementing the stretch attack in Cooja simulator. This type of attack is an extended version of Hello flooding in which rather than sending packets from source to sink, it uses a longer route to cause a higher rate of energy usage throughout the network. Thus, smart sensors should be exhausted earlier than during a Hello flooding attack. In order to compare power consumption, we used three different battery types; 2xAAA (1000 mAh), CR2032 (225 mAh), and CR123A (1500 mAh). This was achieved with the help of Kinetic Battery module of ContikiOS. The network used in this scenario consisted of 9 honest nodes, one malicious node and one server.

\begin{table*}[htb]
    \centering
    \tbl{Average network battery consumption for different battery--attack combinations}
    {\begin{tabular}{@{}cl@{\qquad}r@{\qquad}r@{}}
         \toprule
          &  & \multicolumn{2}{c}{\textbf{Scenario}} \\ 
         \cmidrule(){3-4}
          &  & No Attack & Stretch Attack \\ 
         \toprule
         \multirow{3}{*}{\rotatebox[origin=c]{90}{\textbf{Battery}}} & 2xAAA (1000 mAh) & 11.32\% & 27.22\%\\
         \cmidrule(){2-4} 
          & CR2032 (225 mAh) & 50.40\% & Empty \\ 
         \cmidrule(){2-4} 
          & CR123A (1500 mAh) & 7.55\% & 18.20\% \\ 
        \bottomrule
    \end{tabular}}
    \label{tab:batterycons}
\end{table*}

Table \ref{tab:batterycons} depicts the average battery consumption over the entire network of honest nodes when they are idle and when they are under stretch attack. The duration of the experiment was one hour. If the results for each individual node are taken into account, some have a much higher battery usage than others. Generally, average battery consumption is high when stretch attack occurs. However, the only time the average battery consumption was enough to deplete is when the battery had 225 mAh during the stretch attack.
Using the smallest battery (225 mAh), some nodes started to run out after only 30 minutes, and after 50 minutes six of the nine nodes had completely run out of battery. After an hour, nodes with 1000 mAh battery had been left at only 57\% battery life. Although the stretch attack did not deplete any of the nodes using the 2xAAA battery or the CR123A battery, a rogue node emitting too many Hello packets could achieve that.

\subsection{Power consumption} 
A comparison between power consumption for each implemented attack is presented in Table \ref{tab:compatt}. As is it depicted, both rank attack and selective forwarding have no effect on the overall power consumption. However, the rank attack can be used to make other attacks more severe while the selective forwarding degrades the service of the system by impairing the packet throughput. The `Hello' flooding attack despite multicasting messages, it achieves the lowest increase in power consumption. This is because the malicious node has not been configured to send a large amount of packets to other nodes. Also, the packets sent are small in size because they contain no data and their header is small. Looking at packet flooding attack, it uses twice as much energy as the `Hello' flooding. The packets sent by this are large as they contain all the routing and protocol information as well as data. So, this increases the energy expenditure per packet sent. The most severe attack according to the table is versioning attack. It consumes more than double the power consumption of a packet flooding. This attack can cause serious damage because it affects all network nodes as the malicious node constantly sends messages to all of its neighbours trying to reconstruct the network. 

\begin{table*}[ht]
    \centering
     \tbl{Power consumption with and without attacks}
    {\begin{tabular}{@{}ccccc@{}}
    \toprule
        \textbf{Attack} &
        \textbf{Power consumption (mW)} &
        \textbf{Increase (mW)} & 
        \textbf{Increase (\%)} \\
        \midrule
        No attack & 9.202 & Zero & Zero \\
        \midrule
        Packet flooding & 44.726 & 35.524 & 386\% \\
        \midrule
        Selective forwarding & 9.729 & Negligible & Negligible \\
        \midrule
        Versioning & 105.834 & 96.632 & 1050\% \\
        \midrule
        Rank & 9.125 & Negligible & Negligible \\
        \midrule
        Hello flooding & 24.945 & 15.743 & 171\% \\
    \bottomrule 
    \end{tabular}}
    \label{tab:compatt}
\end{table*}

\section{Comparison between various attacks} \label{comparison}
In this section we compare all the attacks we considered with respect to the various resource consummation criteria. The results are compiled in Table~\ref{tab:compallcrit} to facilitate comparison. As the table shows, Versioning Attacks are quite powerful in wasting the resources of the network, achieving around 4 times the increase in CPU time and TX time, more than around 4 times the decrease in LPM time, and around 3 times the increase in battery consumption compared to those of Packet Flooding and Hello Flooding attacks, respectively. At the same time, Versioning Attacks inflict the most RX time increase, with Packet Flooding and Hello Flooding following suite. Packet Flooding and Hello Flooding perform similarly with respect to CPU time, LPM time, and TX time. However, Packet Flooding is able to waste almost four times the amount of RX time and double the battery power compared to Hello Flooding. Selective Forwarding and Rank Attacks do not waste much of CPU time, RX/TX time, or battery but they affect the network in other ways as discussed before. 

\begin{table*}[ht]
    \centering
     \tbl{Comparison of various attacks with respect to usage of various resources}
    {\begin{tabular}{@{}cccccc@{}}
    \toprule
        \textbf{Attack} &
        \textbf{CPU Time} &
        \textbf{LPM} & 
        \textbf{TX Time} &
        \textbf{RX Time} &
        \textbf{Battery Consumption} \\ 
        \textbf{Type} &
        \textbf{Increase} &
        \textbf{Decrease} & 
        \textbf{Increase} &
        \textbf{Increase} &
        \textbf{Increase} \\ 
    \midrule
        Packet Flooding & 265\% & 3\% & 1279\% & 329\% & 386\% \\ 
    \midrule     
        Hello Flooding & 226\% & 2\% & 1275\% & 81\% & 171\% \\ 
    \midrule     
        Selective Forwarding & negligible & negligible & negligible & negligible & negligible \\ 
    \midrule     
        Versioning Attack & 1146\% & 12\% & 4750\% & 478\% & 1050\% \\ 
    \midrule     
        Rank Attack & negligible & negligible & negligible & negligible & negligible \\ 
    \bottomrule 
    \end{tabular}}
    \label{tab:compallcrit}
\end{table*}

\section{Conclusion} \label{conclusion}
In this paper, several aspects of smart devices and how different network setups affect their battery life is studied. Literature showed that network edge devices are insecure and specifically their hardware components, such as batteries and connectivity capabilities, could be easily exploited. Following this, we explored five main attacks that can be launched to exhaust devices' small batteries. Our results show that certain attacks are able to dramatically increase energy consumption and affect device's operations. Furthermore, a scenario implementing the stretch attack was done in order to study the effects of this extreme case. Simulations showed that when sending even small numbers of packets, typical IoT devices' batteries can easily be drained in 60 minutes. Moreover, devices closer to rogue nodes are more affected by the attacks than others.

In our future work we care planning to compare the derived simulation results to the results obtained via actual IoT hardware implementations. Finally, we will investigate the impact on the battery life when combining two or more attack types.

\section*{Disclosure Statement}
No potential conflict of interest was reported by the authors.

\bibliographystyle{unsrt}
\bibliography{MyBib}

\begin{thebibliography}{10}

\bibitem{kanan2016iot}
Riad Kanan.
\newblock {IoT devices: The quest for energy security}.
\newblock In {\em IEEE 59th International Midwest Symposium on Circuits and
  Systems (MWSCAS)}, pages 1--4, Abu Dhabi, United Arab Emirates, 2016.

\bibitem{ioulianou2019battery}
Philokypros~P Ioulianou, Vassilios~G Vassilakis, and Michael~D Logothetis.
\newblock Battery drain denial-of-service attacks and defenses in the internet
  of things.
\newblock {\em Journal of Telecommunications and Information Technology}, 2019.

\bibitem{gigli2011internet}
Matthew Gigli and Simon~GM Koo.
\newblock Internet of {T}hings: Services and applications categorization.
\newblock {\em Adv. Internet of Things}, 1(2):27--31, 2011.

\bibitem{cisco_2014}
CISCO.
\newblock {The Internet of Things Reference Model}, 2014.

\bibitem{mosenia2017comprehensive}
Arsalan Mosenia and Niraj~K Jha.
\newblock A comprehensive study of security of {Internet-of-Things}.
\newblock {\em IEEE Transactions on Emerging Topics in Computing},
  5(4):586--602, 2017.

\bibitem{jardosh2005understanding}
Amit~P Jardosh, Krishna~N Ramachandran, Kevin~C Almeroth, and Elizabeth~M
  Belding-Royer.
\newblock Understanding congestion in ieee 802.11b wireless networks.
\newblock In {\em Proceedings of the 5th ACM SIGCOMM conference on Internet
  Measurement}, pages 25--25, Berkeley, USA, 2005.

\bibitem{Wi-FiAlliance}
Wi-Fi Alliance.
\newblock {Wi-Fi HaLow}.

\bibitem{deng2014ieee}
Der-Jiunn Deng, Kwang-Cheng Chen, and Rung-Shiang Cheng.
\newblock {IEEE} 802.11 ax: Next generation wireless local area networks.
\newblock In {\em 10th International Conference on Heterogeneous Networking for
  Quality, Reliability, Security and Robustness (QShine)}, pages 77--82,
  Rhodes, Greece, 2014. IEEE.

\bibitem{ray2016bluetooth}
Partha~Pratim Ray and Sneha Agarwal.
\newblock Bluetooth 5 and {Internet of Things}: Potential and architecture.
\newblock In {\em International Conference on Signal Processing, Communication,
  Power and Embedded System (SCOPES)}, pages 1461--1465, Paralakhemundi, India,
  2016. IEEE.

\bibitem{lansford2001wi}
Jim Lansford, Adrian Stephens, and Ron Nevo.
\newblock {Wi-Fi} (802.11 b) and bluetooth: Enabling coexistence.
\newblock {\em IEEE Network}, 15(5):20--27, 2001.

\bibitem{rfc6550}
Winter T, Thubert P, Brandt A, Hui J, et~al.
\newblock Rpl: Ipv6 routing protocol for low-power and lossy networks.
\newblock {\em RFC}, 6550:1--157, 2012.

\bibitem{trickle1}
P.~Levis, T.~Clausen, J.~Hui, O.~Gnawali, and J.~Ko.
\newblock {The Trickle algorithm}.
\newblock \url{https://www.rfc-editor.org/rfc/rfc6206.txt}, 2011.

\bibitem{google}
Google.
\newblock {Overview of Internet of Things}.
\newblock \url{https://cloud.google.com/solutions/iot-overview}.

\bibitem{voica_2016}
Alex Voica.
\newblock A guide to internet of things ({IoT}) processors, Jun 2016.

\bibitem{brooks2017ultra}
David Brooks and John Sartori.
\newblock Ultra-low-power processors.
\newblock {\em IEEE Micro}, 37(6):16--19, 2017.

\bibitem{batterysize}
Silicon Labs.
\newblock Battery size matters.
\newblock
  \url{https://www.silabs.com/documents/referenced/white-papers/battery-life-in-connected-wireless-iot-devices.pdf}.

\bibitem{calhoun2015ultra}
Benton~H Calhoun and David~D Wentzloff.
\newblock Ultra-low power wireless socs enabling a batteryless iot.
\newblock In {\em Hot Chips Symposium}, pages 1--45, Cupertino, USA, 2015.

\bibitem{zhao2013survey}
Kai Zhao and Lina Ge.
\newblock A survey on the {I}nternet of things security.
\newblock In {\em 9th International Conference on Computational Intelligence
  and Security (CIS)}, pages 663--667, Leshan, China, 2013. IEEE.

\bibitem{secintel}
Security for intelligent, connected {IoT} edge nodes.
\newblock
  \url{http://ww1.microchip.com/downloads/en/DeviceDoc/Atmel-8994-Security-for-Intelligent-Connected-IoT\-Edge-Nodes_Whitepaper.pdf}.

\bibitem{krentz2017countering}
Konrad-Felix Krentz, Christoph Meinel, and Hendrik Graupner.
\newblock Countering three denial-of-sleep attacks on {ContikiMAC}.
\newblock In {\em Proceedings of the International Conference on Embedded
  Wireless Systems and Networks (EWSN 2017)}, pages 108--119, Uppsala, Sweden,
  2017.

\bibitem{raymond2009effects}
David~R Raymond, Randy~C Marchany, Michael~I Brownfield, and Scott~F Midkiff.
\newblock Effects of denial-of-sleep attacks on wireless sensor network {MAC}
  protocols.
\newblock {\em IEEE Transactions on Vehicular Technology}, 58(1):367--380,
  2009.

\bibitem{thakur2013introduction}
Neha Thakur and Aruna Sankaralingam.
\newblock Introduction to jamming attacks and prevention techniques using
  honeypots in wireless networks.
\newblock {\em International Journal of Computer Science and Information
  Technology \& Security}, 3(2):202--207.

\bibitem{singh2010hello}
Virendra~Pal Singh, Sweta Jain, and Jyoti Singhai.
\newblock Hello flood attack and its countermeasures in wireless sensor
  networks.
\newblock {\em International Journal of Computer Science Issues (IJCSI)},
  7(3):23, 2010.

\bibitem{vasserman2013vampire}
Eugene~Y Vasserman and Nicholas Hopper.
\newblock Vampire attacks: draining life from wireless ad hoc sensor networks.
\newblock {\em IEEE Transactions on Mobile Computing}, 12(2):318--332, 2013.

\bibitem{umakanth2013detection}
B~Umakanth and J~Damodhar.
\newblock Detection of energy draining attack using {EWMA} in wireless ad hoc
  sensor networks.
\newblock {\em International Journal of Engineering Trends and Technology
  (IJETT)}, 4:3691--3695.

\bibitem{rathi2017wearable}
Neeraj Rathi, Monika Kakani, Mohamed El-Sharkawy, and Maher Rizkalla.
\newblock Wearable low power pre-fall detection system with {IoT} and bluetooth
  capabilities.
\newblock In {\em IEEE National Aerospace and Electronics Conference (NAECON)},
  pages 241--244, Dayton, USA, 2017.

\bibitem{mayzaud2014study}
Anth{\'e}a Mayzaud, Anuj Sehgal, R{\'e}mi Badonnel, Isabelle Chrisment, and
  J{\"u}rgen Sch{\"o}nw{\"a}lder.
\newblock A study of {RPL DODAG} version attacks.
\newblock In {\em IFIP International Conference on Autonomous Infrastructure,
  Management and Security}, pages 92--104, Brno, Czech Republic, 2014.
  Springer.

\bibitem{wallgren2013routing}
Linus Wallgren, Shahid Raza, and Thiemo Voigt.
\newblock {Routing attacks and countermeasures in the RPL-based Internet of
  Things}.
\newblock {\em International Journal of Distributed Sensor Networks},
  9(8):794326, 2013.

\bibitem{lahmar2012wireless}
Khawla Lahmar, Rym Ch{\'e}our, and Mohamed Abid.
\newblock Wireless sensor networks: Trends, power consumption and simulators.
\newblock In {\em Sixth Asia Modelling Symposium (AMS)}, pages 200--204, Bali,
  Indonesia, 2012. IEEE.

\bibitem{contiki}
Contiki: The {O}pen {S}ource {O}{S} for the {I}nternet of {T}hings.
\newblock \url{http://www.contiki-os.org/}.
\newblock Accessed: 2019-01-25.

\bibitem{jan2013denial}
Mian~Ahmad Jan and Muhammad Khan.
\newblock Denial of service attacks and their countermeasures in wsn.
\newblock {\em IRACST--International Journal of Computer Networks and Wireless
  Communications (IJCNWC)}, 3:1--6, 2013.

\bibitem{d'hondt_bahmad_vanhee_2016}
Alexandre D'Hondt, Hussein Bahmad, and Jeremy Vanhee.
\newblock {RPL} attacks framework.
\newblock
  \url{https://github.com/dhondta/rpl-attacks/blob/master/doc/report.pdf},
  2016.

\bibitem{zolertiaz1}
Zolertia z1 datasheet.
\newblock
  \url{https://github.com/Zolertia/Resources/blob/master/Z1/Hardware/Revision
  C/Datasheets/Zolertia Z1 datasheet Revision C.pdf}, 2010.

\bibitem{alkaline_handbook}
Panasonic Corporation.
\newblock Alkaline handbook.
\newblock
  \url{https://eu.industrial.panasonic.com/sites/default/pidseu/files/downloads/files/id_alkaline_1203_e.pdf}.

\end{thebibliography}

\vspace{2\baselineskip}

\par
\textbf{Ryan Smith}{ is a fifth year student completing his integrated masters in Computer Science at the University of York, UK. For his masters project he is currently investigating the application of different meta-heuristic techniques for the task mapping problem on network-on-chips. His research interests include nature-inspired computing, cryptography, IoT (Internet of Things) and wireless sensor networks.}

\vspace{2\baselineskip}
\par

\textbf{Daniel Palin}{ has recently completed his Master of Science in Cyber Security at the University of York. His research interests are in the areas of IoT security and DoS attacks mitigation.}

\vspace{2\baselineskip}
\par
\textbf{Philokypros P. Ioulianou} {received his B.Sc. degree in Computer Science from the University of Cyprus in 2016, and MSc in Advanced Computer Science with specialization in Computer Security from the University of Manchester in 2017. He is currently a PhD student at the University of York, UK. His research interests are in the area of Computer and Network Security, IoT (Internet of Things) and Wireless Sensor Security.}

\vspace{2\baselineskip}
\par
\textbf{Vassilios G. Vassilakis}{ is a Lecturer (Assistant Professor) in Cyber Security at the University of York, UK. He received his Ph.D. degree in Electrical and Computer Engineering from the University of Patras, Greece in 2011.
From 2011 to 2013, he was with the Network Convergence Laboratory, University of Essex, where he conducted research on information-centric networking. In 2013, he joined the Institute for Communication Systems, University of Surrey, and conducted research on 5G networks. After that, he was with the Computer Laboratory, University of Cambridge, where he conducted research on future Internet technologies. In 2015, he joined the Secure and Dependable Software Systems Research Cluster at the University of Brighton, where he conducted research on 5G network security and contributed to the EC H2020 SESAME project. 
He\rq{s} been involved in EU, UK, and industry funded R\&D projects related to the design and analysis of future mobile networks and Internet technologies. 
His main research interests are in the areas of network security, next-generation wireless and mobile networks, Internet of things, and software-defined networks.
He\rq{s} served as a Guest Editor in Elsevier's Optical Switching \& Networking, IET Networks, and IEICE Transactions on Communications, and for the TPC of IEEE ICC and IEEE GLOBECOM. Contact him at vasileios.vasilakis@york.ac.uk.}

\vspace{2\baselineskip}
\par
\textbf{Siamak F. Shahandashti}{ is a Lecturer (Assistant Professor) at the University of York, UK. His research interests include applied cryptography, privacy-preserving protocols, biometric authentication, and electronic voting. He received his PhD from Wollongong University, Australia, and has been with Victoria University, Sydney, École Normale Supérieure, Paris, and Newcastle  University, UK, before joining York. He regularly serves as the programme committee member of International Conference on Privacy, Security and Trust (PST) and as reviewer to journals such as ACM TISSEC, IEEE TDSC and TIFS, and conferences such as CCS, PKC, and ESORICS. Contact him at siamak.shahandashti@york.ac.uk.}

\end{document}